# The mechanism of multi-anode electrode geometry applied to vacuum arc thruster

Weisheng Cui, Wenzheng Liu*, Yongjie Gao, Xiuyang Chen
School of Electrical Engineering, Beijing Jiaotong University, Beijing, China, 100044

**Abstract**
A multi-anode electrode geometry suitable for vacuum breakdown is proposed in this paper. This electrode geometry forms a unique electric field distribution based on its anode configuration which is different from the conventional electrode geometry. There exists a region A between the insulated-anode and the remote-anode where the electric field vector has opposite directions. The existence of region A directly affects the movement of electrons in the initial stage of the discharge, thereby changing the process of vacuum breakdown. The variation of discharge process affects the generation and propagation mechanism of plasma, forming a plasma plume in favor of the propulsion performance of thrusters. The photograph of plasma plume and the electron density spatial distribution of plasma plume preliminarily prove the pinch effect of multi-anode electrode geometry on the radial diffusion of plasma. In the optimization of remote-anode of multi-anode electrode geometry, the variation of the distance of cathode and remote-anode (short for D) is found to affect the pinch effect. Through the reduction of D, the directional jet effect of the multi-anode electrode geometry is enhanced under the same discharge parameters. Applying it to the VAT (multi-anode VAT) and comparing it with the VAT based on conventional electrode geometry (conventional VAT), it is found that the propulsion performance of multi-anode VAT is superior to conventional VAT, and it is further promoted with the decrease of D. Under the same discharge parameters, the impulse bit and the thrust-to-power ratio of the multi-anode VAT is able to be increased by a factor of 26 and 2.88 respectively compared with the conventional VAT. Theoretical analysis and investigation of propulsion performance validate the potential application advantages of the multi-anode VAT in the field of nanosatellite propulsion.

## 1. Introduction

Vacuum arc thruster (VAT) can be used in the field of nanosatellite propulsion, because of its small quality, low volume and simple structure. The rapid development of nanosatellite technology has put forward higher requirements on the propulsion performance [1-4]. Therefore, the optimization of thruster propulsion effect is of great practical significance.

An important factor determining the propulsion performance of VAT is its plasma generation characteristics, which is greatly influenced by the electrode structure parameters [5-8]. Lun [9] investigated the use of conically shaped convergent cathode surface profiles in a low-power coaxial VAT design, and found that modifying the profile of the cathode surface can affect the plasma jet's plume distribution and even improve thrust production in certain cases. Lukas [10] proposed that at the same power level, thrust was increased by utilizing an ablative anode apart from the conventional erosion of cathode material. However, the VAT in previous studies is based on the conventional electrode geometry (short for conventional VAT). The cathode and the anode are usually arraged inside of the VAT in conventional electrode geometry and the Z-pinch [11,12], which improves the directionality of plasma, only exists inside of the thruster. When the plasma flows out of the nozzle, it tends to diffuse in radial direction and reduce the propulsion performance of the VAT.

The application of external magnetic field is an important method for the optimization of the plasma plume [13-15]. Keidar [16] investigated the magnetically enhanced coaxial VAT and the VAT with ring electrodes and found that the radial expansion of the plasma plume is significantly decreased with the magnetic field under typical conditions. Except for the commonly used axial magnetic field, Zhuang [17] proposed to use a curved magnetic field to obtain efficient plasma transport and high directional plume. It was demonstrated that the applied magnetic field can enhance the plasma plume and lead to homogenous erosion critical for operational lifespan of thruster. However, the use of external magnetic usually needs extra power supply and control circuit, which will add the quality and volume of the thruster and does not contribute to the application advantage of VATs.

In this paper, a discharge method of multi-anode electrode geometry is proposed based on the conventional electrode geometry, achieving a long gap vacuum breakdown. This method changes the discharge characteristics of electrodes through the specially designed multiple anodes, thus changing the generation and propagation characteristics of vacuum metal plasma. It is able to direct the arc current to pass through the plasma plume, thereby extending the Z-pinch from being limited within the thruster in conventional VAT to the plasma plume outside the thruster nozzle. In addition, it does not need to introduce any external magnetic field, so it eliminates the problems caused by the application of external magnetic field. The reminder of this paper is organized as follows: the experimental system is introduced in the second part. The VAT based on the multi-anode electrode geometry (short for multi-anode VAT) is introduced along with its comparison with the conventional VAT in the third part. This part elaborates the discharge characteristics of the multi-anode electrode geometry in detail and introduces the mechanism of this Z-pinch. The fourth part investigates the influence of parameters of the multi-anode electrode geometry on the discharge characteristics and the pinch effect. The fifth part validates the improvements of propulsion performance of the multi-anode VAT and it is followed by the conclusion and prospect.

## 2. Experimental setup

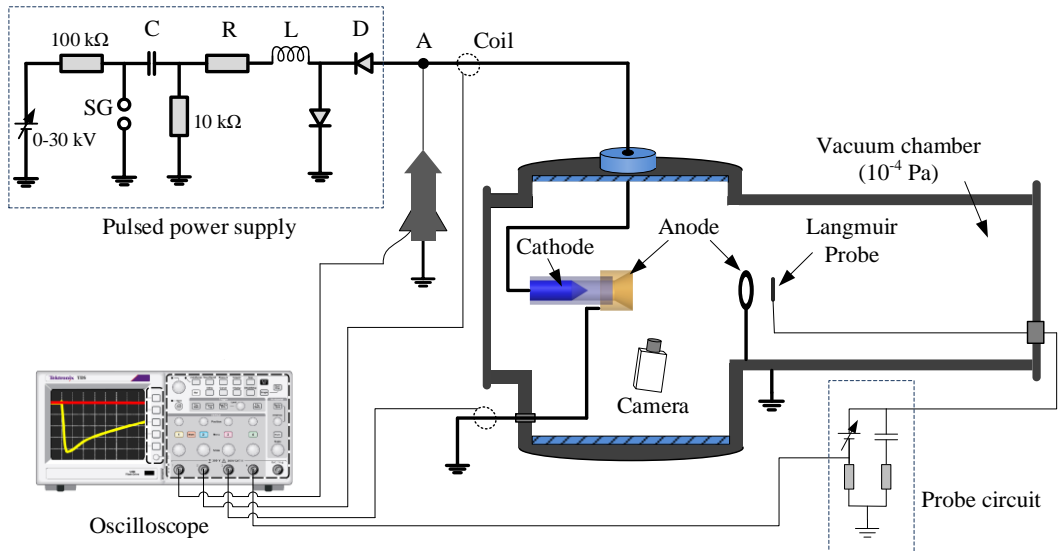

Figure 1. Schematic diagram of the experimental system.

The experimental system is shown in figure 1. It includes a pulsed power supply, a vacuum discharge device and diagnostic systems. The discharge process is as follows: the energy storage capacitance C (0.1 μF) is charged to a high voltage before the discharge. The spherical gap SG is controlled by a pulse signal to trigger the discharge. When SG is closed, the voltage of C will be applied between the cathode and anode through the current-limiting resistance R (27 Ω) and the inductance L (220 μH). Therefore, the discharge is produced due to the vacuum breakdown. The L is designed to extend the duration time of arc current so as to increase the generation of plasma. A diode D is set in series with the cathode to prevent the current from flowing back and avoid the occurrence of current oscillation. The output of this pulsed power supply is a negative voltage with a maximum amplitude of 30 kV. The vacuum chamber can be pumped down to the pressure of $10^{-4}$ Pa, ensuring that the discharge happens in a high vacuum environment. During the discharge, the interelectrode voltage (voltage between point A and the ground) is measured by a high voltage probe (TEK-P6015A); the currents flowing through the cathode and the anode are obtained by two Rogowski coils. The plasma electron density is measured by an improved Langmuir probe method [6,18-20], and the discharge phenomenon is captured by a camera.

## 3. The operational principle of multi-anode VAT

### 3.1. Discharge characteristics

The VAT studied in this research is based on coaxial electrode geometry. The conventional coaxial electrode geometry can be seen in figure 2(a), which mainly consists of a cathode, an anode and an insulator. In this electrode geometry, a solid polytetrafluoroethylene (PTFE) tube is applied as the insulator, and its inner diameter and outer diameter are 5 mm and 7 mm respectively. The cathode is made from lead and the anode stainless steel: the cathode is cylindrical (diameter of 5 mm) with one end of a cone (cone angle of 60°) and is placed in the PTFE tube; the anode is in the shape of a trumpet nozzle and is fixed to one end of the PTFE tube. The axial distance between the cathode cone tip and the end of the PTFE tube is 6 mm. The multi-anode electrode geometry (it can be seen in figure 2(b) and figure 3) is designed based on the conventional coaxial electrode geometry. In the multi-anode electrode geometry, the first anode next to the cathode is insulated completely (insulated-anode); the second anode is in the shape of a ring (remote-anode) and is placed 100 mm axial direction away from the cathode. The remote-anode has an inner diameter of 5 mm and an outer diameter of 15 mm.

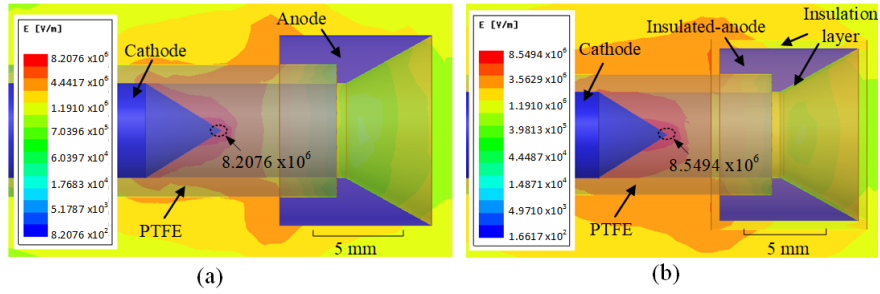

Figure 2. The electric field intensity distribution of the conventional electrode geometry (a) and the multi-anode electrode geometry (b) (the remote-anode is not displayed in the graph).

The VATs in this research utilize the mechanism of field emission to produce the discharge and form vacuum metal arc between the cathode and the anode. When a high voltage is applied between the cathode and the anode, a strong electric field will be formed on the cathode which will induce field emission and thereby producing the vacuum breakdown [21]. The ANSYS Maxwell 3D simulation software is used to analyze the electric field distribution of this electrode geometry when a voltage of 10 kV is applied between the cathode and the anode. The electric field intensity distribution for the two electrode geometries is shown in figure 2.

It has been seen from figure 2 that the strongest electric field appears on the cathode cone tip for both electrode geometries, reaching the level of $8\times 10^6$ V/m. The emitters (mostly some protrusions with diameter of less than 10 μm) on the cathode tip surface usually increase the effective electric field intensity by a factor of dozens to hundreds due to a field enhancement factor β [22]. Therefore, it is possible for the emitters to reach the critical field intensity for the onset of electron field emission ($10^8$ V/m) [23] at this voltage level. It has been known that the cathode is able to form the discharge channel with the anode directly in conventional electrode geometry. While, the multi-anode electrode geometry, due to its special arrangement of anodes, has different discharge characteristics.

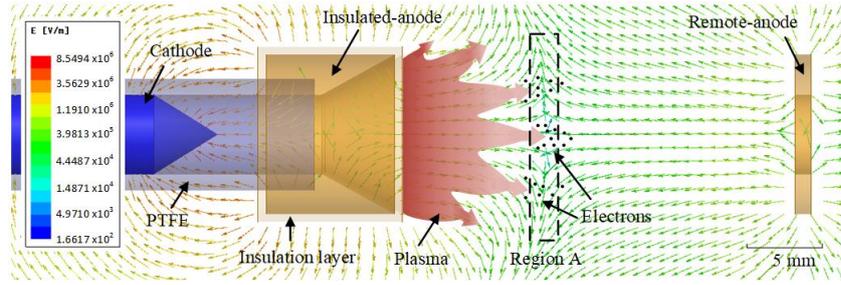

Figure 3. The distribution of electric field vector of the multi-anode electrode geometry, the schematic diagram of the plasma and the extracted electrons are depicted in the graph.

The distribution of electric field vector for the multi-anode electrode geometry is shown in figure 3(the characteristic of electric field vector distribution remains unchanged when the distance between the cathode and the remote-anode decreases from 100 mm to 40 mm; therefore, the distance is set to 40 mm for display and analysis.) In the figure, there is a region (called region A) between the insulated-anode and the remote-anode where the electric field vector has opposite directions. It has been found in further simulation that the region A remains unchanged before the arrival of the plasma. And the region A moves forward together with the front edge of the plasma after the plasma arrives at its position and propagates towards the remote-anode. The existence of region A has an important effect on the directional drift of electrons in the initial stage of discharge.

It is generally accepted that the directional drift of electrons is determined by the electric field and the directional propagation of metal plasma is mainly affected by the metal ions in the pulse discharge [24]. Therefore, it is concluded that the electrons emitted from the cathode mainly move towards the insulated-anode under the influence of electric field in the initial stage of discharge. Some of the electrons might pass through the insulated-anode and move towards the remote-anode. However, the directional drift of electrons is suppressed due to the influence of electric field between the insulated-anode and the region A and they can hardly reach the remote-anode. Since the cathode cannot be connected to both the insulated-anode and the remote-anode, the drop of the interelectrode voltage is suppressed and the generation of the plasma is sustained. When the plasma arrives at the region A and propagates towards the remote-anode (which is indicated in the graph), some electrons in the front edge of the plasma could be extracted in region A and reach the remote-anode under the influence of the electric field between the region A and the remote-anode. Eventually, the arc will be formed when the metal plasma arrives at the remote-anode.

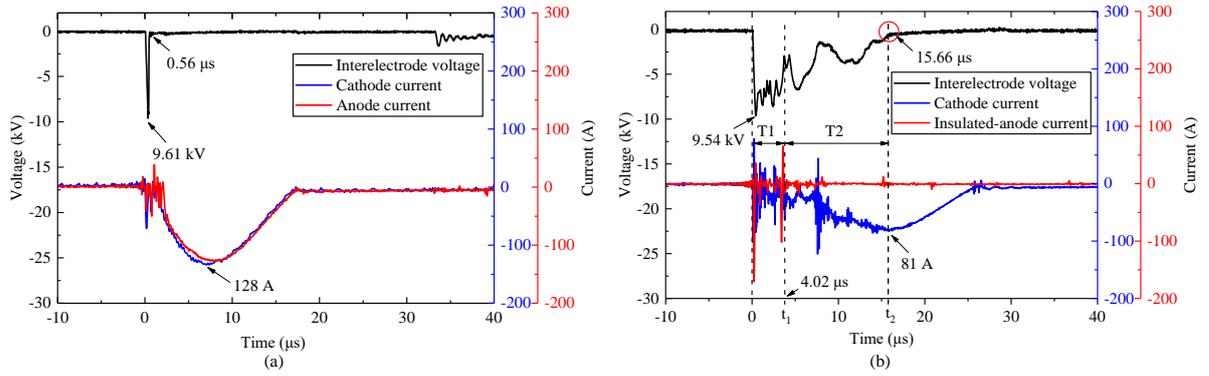

Figure 4. Interelectrode voltage and current of the conventional electrode geometry (a) and the multi-anode electrode geometry (b).

A series of experiments have been carried out and the typical waveforms of the interelectrode voltage and discharge current of the two electrode geometries with the same discharge parameters are shown in figure 4. It has been seen from figure 4(a) that the breakdown of the conventional electrode geometry occurs instantaneously and the interelectrode voltage decreases to arc voltage (the arc voltage is below 200 V and not displayed in the graph) within 0.56 μs. The arc current flowing through the cathode and the anode (short for cathode current and anode current, respectively) increases synchronously after the vacuum breakdown. The metal plasma is ejected from the nozzle and forms a plasma plume. According to the Kirchhoff current law, the current flows through the plasma plume is tiny.

The multi-anode electrode geometry has been found to have the same discharge reliability as the conventional electrode geometry through thousands of discharges. The breakdown voltage of multi-anode electrode geometry is approximately the same with that of the conventional electrode geometry. While, the interelectrode voltage and the discharge current of the multi-anode electrode geometry has different characteristics during the discharge. It has been seen from figure 4(b) that the current flowing through the insulated-anode (short for insulated-anode current) is basically 0 during the discharge and the amplitude of cathode current reduces to 58.60% of that of the conventional electrode geometry. The interelectrode voltage and the cathode current show different characteristics in different time intervals. In the first time interval of T1 (0-t1), the interelectrode voltage and the cathode current do not change obviously but exhibit some fluctuations after the discharge. It is caused by the fact that the field emission electrons cannot reach the remote-anode instantly in the initial stage of the discharge. The interelectrode voltage starts to drop and the cathode current begins to grow with fluctuations in the waveforms in the second time interval of T2 (t1-t2). In this period the plasma arrives at the region A and the region A moves forward with the front edge of the plasma. An increasing number of electrons is extracted and flows into remote-anode under the influence of the electric field, which leads to the change of the interelectrode voltage and the discharge current. At the end of T2, the plasma reaches the remote-anode and the vacuum arc is formed. Therefore, the interelectrode voltage drops to the arc voltage and the fluctuations in the current waveform is eliminated. The variation trend of interelectrode voltage and the cathode current is able to verify the analysis of discharge process concluded above.

### 3.2. Plasma plume

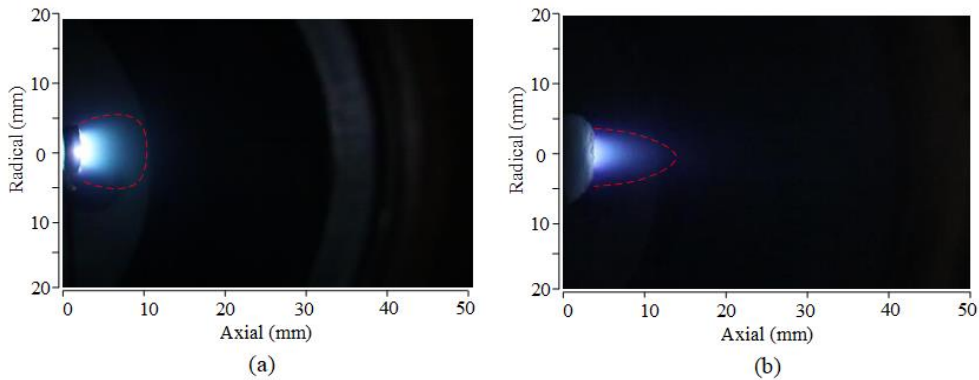

Figure 5. The plasma plume of the conventional VAT (a) and the multi-anode VAT (b). The remote-anode is not shown in the graph (b) because of its long distance with the cathode.

It is inferred from the discharge characteristics that the plasma plume formed in the multi-anode VAT can differ from that of the conventional VAT. The photographs of the plasma plume for both VATs are captured and shown in figure 5. It has been observed from figure 5(a) that the plasma plume of the conventional VAT is bright and divergent outside of the anode nozzle. It has been known that because of the large current, the generation of plasma is fierce and the density of charged particles is increased [24-26]. While, the plasma plume exhibits an obvious diffusion in radial direction. In contrast, the plasma plume of the multi-anode VAT (figure 5(b)) exhibits less brightness but more distinct convergence effect. It is inferred that the amount of the generated plasma in multi-anode VAT is smaller with the decrease of arc current. While, the plasma plume is constricted and enhanced by the current sheet and has a notable directionality. The differences of operation principle between the multi-anode VAT and the conventional VAT are depicted and shown in figure 6.

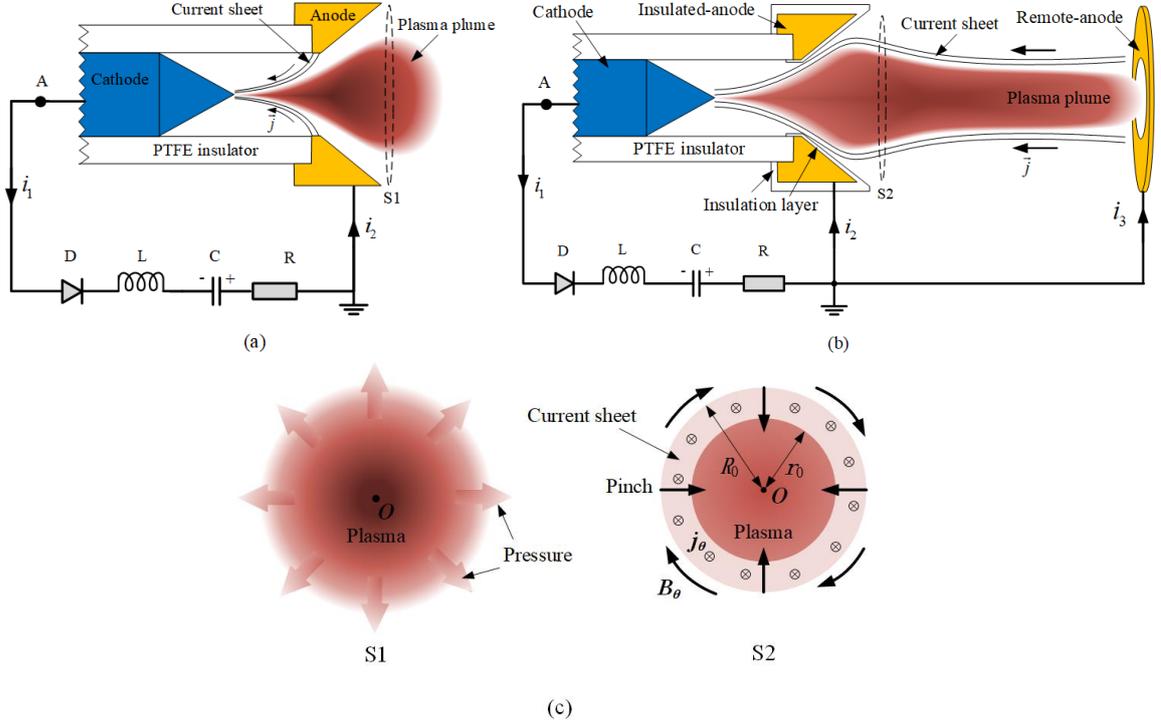

Figure 6. The schematic diagrams demonstrating the principle of the conventional VAT (a) and the multi-anode VAT (b); the schematic diagram of plasma plume cross profile of conventional VAT (S1) and multi-anode VAT (S2) (c).

As shown in figure 6(a), the current sheet is formed within the thruster for the conventional VAT. The plasma plume outside of the nozzle diffuses due to high thermodynamic pressure in the plasma. By comparison, it is found from figure 6(b) that the arc current flows through the plasma plume between the cathode and the remote-anode for the multi-anode VAT. And the plasma is constricted by the current sheet completely in its propagation process. The differences of the plasma plume cross profiles between the two VATs are shown in figure 6(c). The plasma plume of the conventional VAT is not constricted and shows gradient change in radial direction. However, the plasma plume of the multi-anode VAT is constricted by the current sheet, and the plasma inside of the current sheet is in a state of equilibrium.

The theoretical model of the plasma internal pressure as a function of the radial displacement is established. The plasma, as a fluid consists of charged particles, can be described by the theory of magnetohydrodynamics. According to the previous research [27], the relationship between the pressure $p_\theta$ and radial displacement $r$ is calculated as follows,

$$p_\theta = \begin{cases} \dfrac{\mu_0 i^2}{8\pi^2 R_0^2}, & r \leq r_0 \\ \dfrac{\mu_0 i^2}{8\pi^2} \dfrac{R_0^4 r^2 + r_0^4 r^2 - R_0^2 \left(r_0^4 + r^4\right)}{R_0^2 \left(R_0^2 - r_0^2\right) r^2}, & r_0 < r \leq R_0 \\ 0, & r > R_0 \end{cases} \quad (1)$$

Where $R_0$ and $r_0$ are the outer radius and the inner radius of the current sheet respectively; $i$ is the total current of the current sheet; $\mu_0$ is the vacuum permeability. It is found from the equation that the pinch force of the current sheet on the plasma increases with the increase of discharge current. The increase of pinch force will lead to the reduction of plasma plume radius and it can be balanced by the internal pressure of the plasma. According to the law of thermodynamics, $p = k\left(n_i T_i + n_e T_e\right)$, the density and temperature of charged particles inside the plasma will be promoted with the increase of the pressure. Therefore, the pinch effect produced by the multi-anode electrode geometry will increase the density and energy of the charged particles, which is crucial to the performance of the thruster.

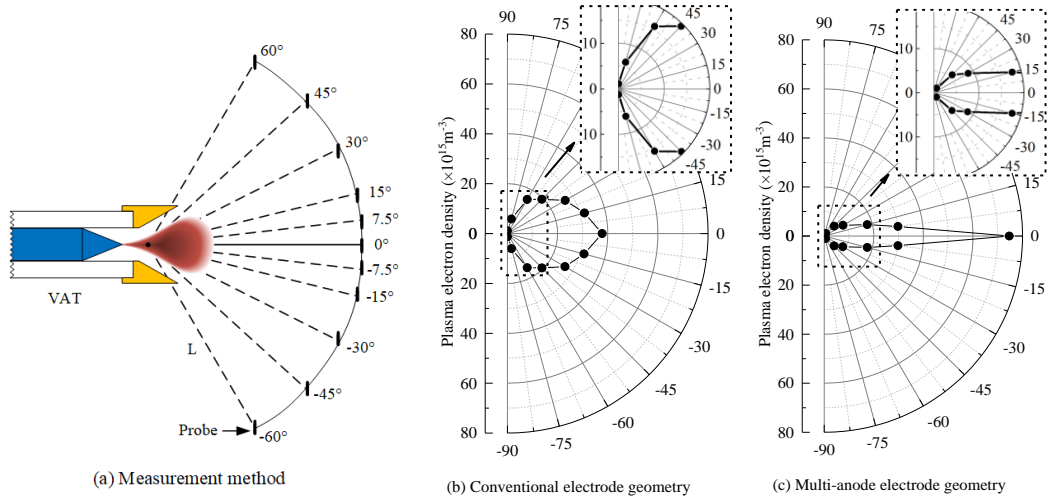

(a) Measurement method  (b) Conventional electrode geometry  (c) Multi-anode electrode geometry

Figure 7. The measurement method of the electron density spatial distribution (a), and the electron density distributions of plasma plume for the conventional VAT (b) and the multi-anode VAT (c).

In order to verify the pinch effect on the plasma, the electron density spatial distribution of the plasma plume for the two VATs with the same discharge parameters are measured. As the plasma is approximately neutral, the electron density is able to reflect the distribution state of plasma in the plasma plume. The specific method is shown in figure 7(a), where the distance L (110 mm) between the VAT and the probe remains constant at different angles. The spatial distribution of electron density is presented in polar coordinate system, where the angle indicates the diffuses direction of plasma and the radial displacement represents the electron density. The electron density distributions for the two VATs are shown in figure 7(b) and (c), respectively.

It has been seen from figure 7 that the electron density distribution of both plasma plumes exhibits a maximum in axial direction (0°) and smaller with the increase of the angle. While, the plasma plume of the multi-anode VAT has a more enhanced electron density in axial direction and reduced electron density in radial direction compared with the conventional VAT. The electron density at 0° for the multi-anode VAT reaches $7.46 \times 10^{16}$ m$^{-3}$, about twice of that for the conventional VAT. The electron density for the multi-anode VAT at other angles decreases obviously: the electron density at 15° and 30° is $1.80 \times 10^{16}$ m$^{-3}$ and $8.65 \times 10^{15}$ m$^{-3}$ respectively, about 24.13% and 11.60% of the density at 0°; in contrast, the electron density for the conventional VAT at 15° and 30° is $3.16 \times 10^{16}$ m$^{-3}$ and $2.65 \times 10^{16}$ m$^{-3}$ respectively, still 83.60% and 70.10% of the density at 0°. The electron density at other angles for the two VATs exhibits huge difference as well.

It has been known that the amount of the plasma generated in the conventional VAT is greater than that of the multi-anode VAT due to the larger arc current. However, without the constriction, the plasma plume of the conventional VAT disperses in radial direction due to the internal pressure, leading to a poor performance of directionality. By contrast, the amount of plasma generated in multi-anode VAT is smaller, but the plasma plume is subjected to the pinch force. The pressure formed by the pinch force increases the density and energy of charged particles in axial direction. As a result, with a smaller amount of plasma, the plasma plume of the multi-anode VAT exhibits a higher electron density in axial direction.

## 4. Parameter optimization of multi-anode electrode geometry

### 4.1. Discharge characteristics of the multi-anode electrode geometry with different D

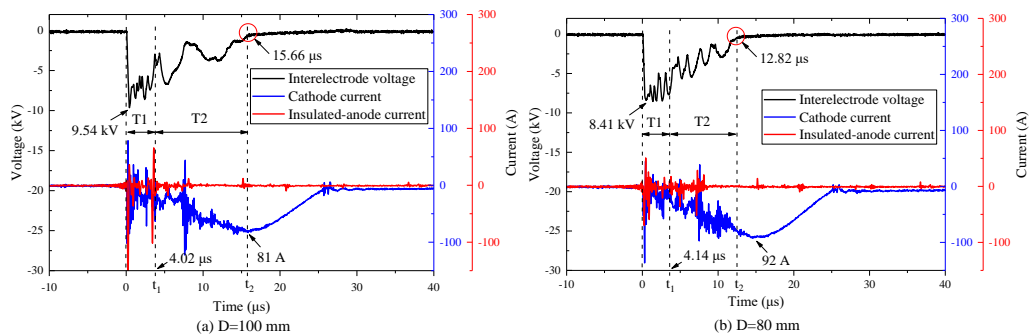

(a) D=100 mm  (b) D=80 mm

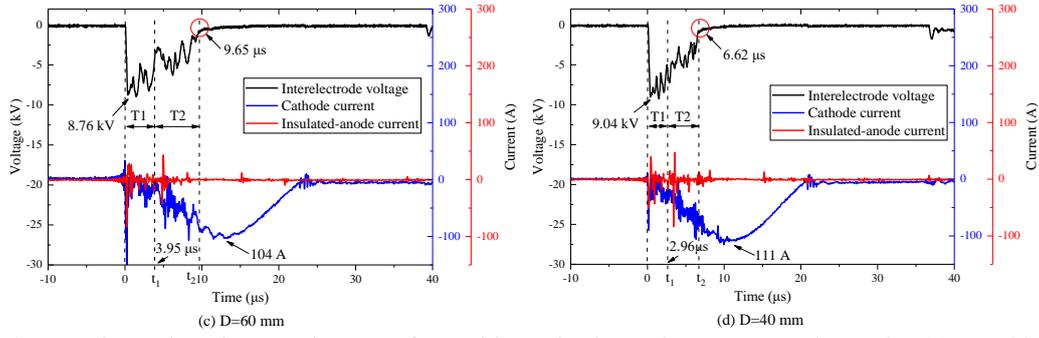

Figure 8. Interelectrode voltage and current for multi-anode electrode geometry when D is 100 mm (a), 80 mm (b), 60 mm (c) and 40 mm (d) respectively.

To investigate the influence of the distance between the cathode and the remote-anode (short for D) on the discharge characteristics of the multi-anode electrode geometry, a series of experiments have been performed with the same discharge parameters. The typical waveforms of interelectrode voltage and arc current when D is 100 mm, 80 mm, 60 mm and 40 mm respectively are shown in figure 8. According to the value of D and the time the interelectrode voltage drops to arc voltage in the graphs, the plasma propagation velocity is calculated to be 6.38 km/s, 6.24 km/s, 6.22 km/s and 6.02 km/s respectively, which are in accord with the propagation velocity of the plasma (about 6 km/s) in the previous research [5].

According to the electric field simulation, when D is 100 mm, 80 mm, 60 mm and 40 mm, the distance between the cathode and the original position of region A is 25.64 mm, 25.81 mm, 23.48 mm, 18 mm, respectively. Assuming the velocity of the plasma is constant during propagation, the duration time of T1 is calculated to be 4.02 μs, 4.14 μs, 3.95 μs and 2.96 μs, respectively. It has been seen from figure 8 that the waveforms of interelectrode voltage and arc current in this time interval are approximately in accord with the characteristics of that in T1. The maintenance of interelectrode voltage is weakened and the cathode current begins to grow in the calculated second time interval (t1 to t2) for each conficuration, which is consistent with the characteristics of that in T2. When the interelectrode voltage drops to the arc voltage, the vacuum arc is formed and the fluctuations in the current waveform is eliminated. The variation of the interelectrode voltage and the discharge current for each configuration conforms to the analysis in the section 3.

In addition, it has been found that the amplitude of the cathode current increases with the decrease of D. The reason is concluded to be that the time required for the plasma to propagate to the remote-anode (T1 plus T2) is reduced with the decrease of D. Therefore, the cathode and the remote-anode is able to form the connection through the generated metal plasma in a shorter time. After the connection, the density of charged particles near the cathode increases rapidly and the vacuum arc is produced. The resistance of the vacuum arc is reduced by the increase of charged particles and the capacitance energy is able to be released quickly through the vacuum arc. Therefore, the reduction of D is able to increase the remaining energy of capacitance the time when the vacuum arc is produced, thereby increasing the amplitude of the vacuum arc during the discharge. The time integral of cathode current is carried out and the total charges passing through the electrodes for each configuration are $1.2 \times 10^{-3}$ C, $1.29 \times 10^{-3}$ C, $1.24 \times 10^{-3}$ C and $1.3 \times 10^{-3}$ C, respectively. Considering the inconsistency of the discharges, the charges passing through the electrodes can be regarded basically the same. The charges passing through the electrodes after the vacuum arc is formed (time of t2) for each configuration are calculated to be $5.04 \times 10^{-4}$ C, $7.51 \times 10^{-4}$ C, $8.99 \times 10^{-4}$ C and $1.04 \times 10^{-3}$ C, respectively. It is consistent with the analysis.

*4.2. The plasma plume of multi-anode VAT with different D*

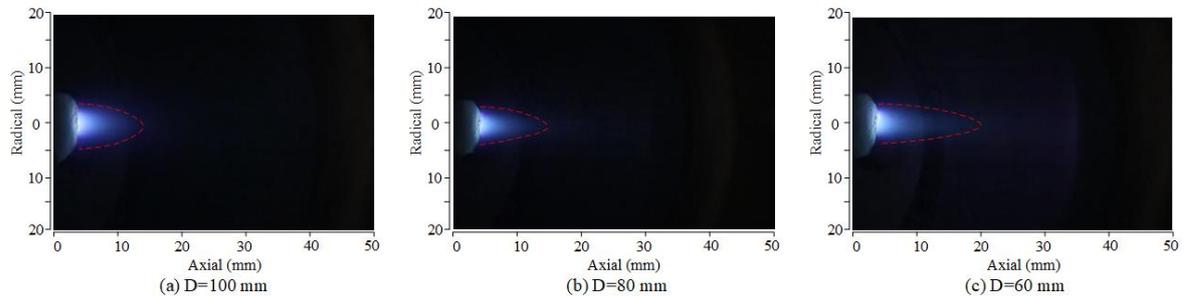

(a) D=100 mm     (b) D=80 mm     (c) D=60 mm

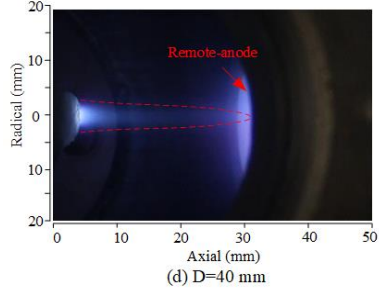

Figure 9. The plasma plume of multi-anode VAT when D is 100 mm (a), 80 mm (b), 60 mm (c) and 40 mm (d) respectively. The remote-anode is not shown in the graph because of its long distance with the cathode except for (d).

The photographs of the plasma plume for each configuration are captured and shown in figure 9. It has been seen from figure 9 that the visible plasma plume is prolonged with the decrease of D. The reason is concluded to be that the decrease of D leads to the increase of arc current, which on one hand promotes the generation of plasma, on the other hand reinforces the pinch force on the plasma and reduces the diffusion of plasma in radial direction. Therefore, the plasma plume is able to maintain a high density at the farther axial distance and the visible length of the plasma plume is prolonged.

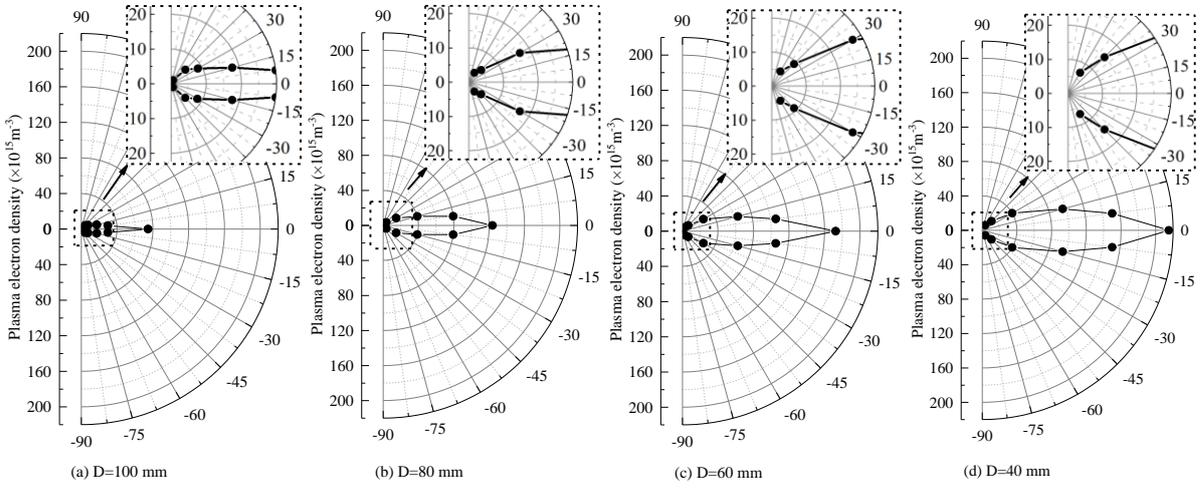

Figure 10. The electron density distribution of plasma plume for the multi-anode VAT when D is 100 mm (a), 80 mm (b), 60 mm (c) and 40 mm (d) respectively.

The electron density spatial distribution of the plasma plume for different D is shown in figure 10. It has been found that the amount of the generated plasma is increased with the decrease of D, which consists with the increase of the arc current. The plasma plumes for each configuration have similar distribution of electron density and exhibit an upward trend with the decrease of D. The electron density at 0° increases from $7.46 \times 10^{16}$ m$^{-3}$ with the D of 100 mm to $2.15 \times 10^{17}$ m$^{-3}$ with the D of 40 mm, increased by a factor of 2.88. However, the percentage of electron density in radial direction to the electron density in axial direction increases with the decrease of D. The percentage of electron density at 15° and 30° increases from 24.13% and 11.60% at D of 100 mm to 44.64% and 18.57% at D of 40 mm.

The reason for this result is concluded as follows: the increase of the amount of the generated plasma will promote the density of charged particles, leading to a rapid increase of the pressure inside of the plasma plume. The increase of current density in current sheet is not large enough to balance this pressure and maintain the same pinch effect. For the reason above, the percentage of the plasma diffusing in radial direction is increased slightly. Despite this fact, the electron density in axial direction is notably increased with the decrease of D. Therefore, the performance of the multi-anode VAT can be promoted.

## 5. The performance of the multi-anode VAT

*5.1. Microthrust measurement system*

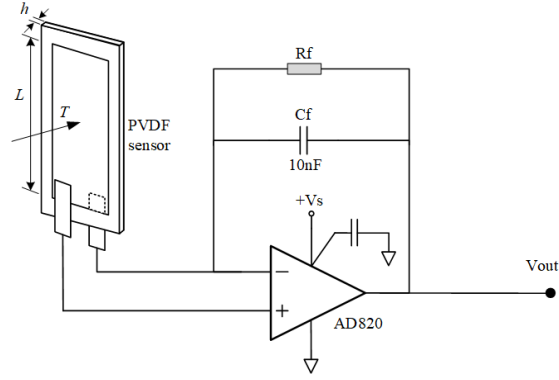

Figure 11. Schematic diagram of the microthrust measurement system.

The performance of the multi-anode VAT is tested in the following experiments. The microthrust measurement system used in the study is proposed by Liu [28], which obtains the thrust of the pulsed electric thruster based on the of Polyvinylidene Fluoride (PVDF) piezoelectric thin-film sensor. The application of piezoelectric thin-film sensor in the thrust measurement of pulsed electric thruster has been calibrated by Wong [29] and it has been confirmed by Takahashi [30] that the force exerted by the plasma flow to the target is in good agreement with the directly measured thrust. The parameters of the measurement system are shown in figure 11. The thrust $T$ is calculated by the following equation,

$$T = \frac{2h^2 C_f}{3K d_{31} L^2} V_{out} \qquad (2)$$

Where $T$ is the microthrust, $h$ is thickness of the PVDF thin-film (28 μm), $L$ is the length of the PVDF thin-film (15 mm), $d_{31}$ is the piezoelectric constant of the PVDF film ($3.3 \times 10^{-11}$ C/N), K is the voltage amplification coefficient.

### 5.2. Measurement of the propulsion performance

In the measurement of the thrust each impulse, the sensor is located at an axial distance of 110 mm from the VAT (the Langmuir probe has been removed before this test). The impulse bit of the VAT is able to be obtained by the integral of the thrust and the duration time. The energy consumed by the VAT is calculated by the waveforms of interelectrode voltage and arc current. The specific equation of the impulse bit and the thrust-to-power ratio are as follows:

$$I_{bit} = \int T dt \qquad (3)$$

$$T/P = \frac{\int T dt}{\int P dt} = \frac{I_{bit}}{E} \qquad (4)$$

Where $I_{bit}$ is the impulse bit, $T/P$ is the thrust-to-power ratio, $P$ is the discharge power of the electrodes, $E$ is the energy consumed by the electrodes each impulse, $t$ is the duration time of the impulse. Meanwhile, to investigate the variation of the performance of VAT with different discharge power, the energy of capacitance C in pulsed power supply is changed to alter the discharge current of the electrodes in the experiment. The impulse bit and thrust-to-power ratio of the VATs as a function of the energy of capacitance C are shown in figure 12.

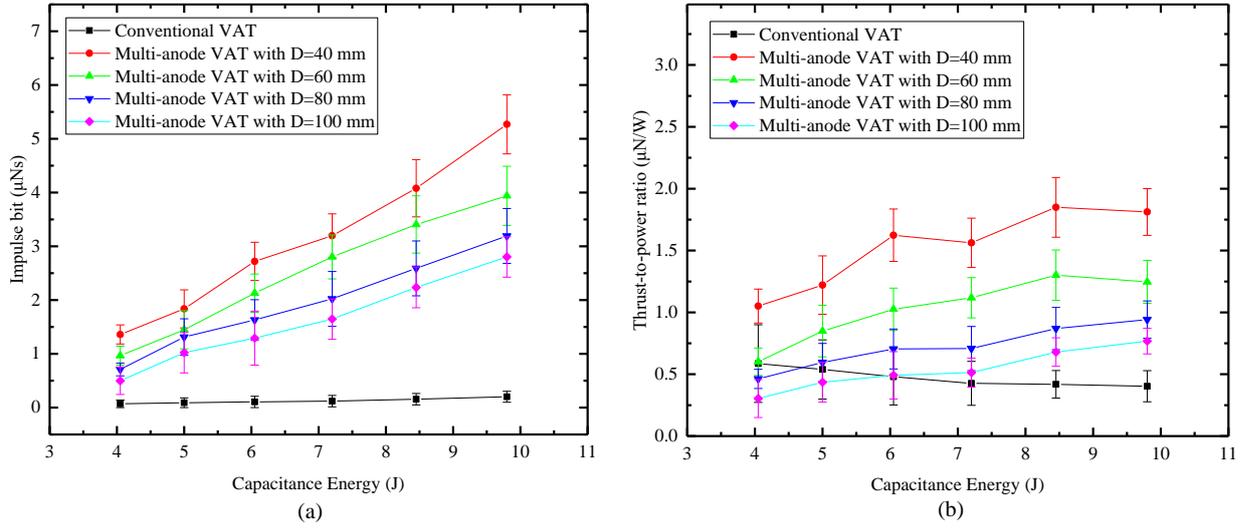

Figure 12. The impulse bit (a) and thrust-to-power ratio (b) for VATs as a function of capacitance energy.

It has been seen from figure 12(a) that the impulse bit of the multi-anode VATs is significantly increased compared with that of the conventional VAT. And the impulse bit of the multi-anode VAT increases with the decrease of D in the same discharge power. The impulse bit of the conventional VAT increases from 0.07 μNs at capacitance energy of 4 J to 0.20 μNs at 9.8 J. In contrast, the impulse bit of the multi-anode VAT with D of 40 mm increases from 1.36 μNs to 5.27 μNs in the same condition. The maximum impulse bit of the multi-anode VAT increases by a factor of 26 compared with that of the conventional VAT. It verifies that the multi-anode electrode geometry proposed in this research is able to effectively promote the propulsion performance of the VAT. Meanwhile, it has been seen from the figure that the promotion is more remarkable with the increase of the discharge power. It has been known from section 4 that the density of the plasma plume in axial direction for the multi-anode VAT increases with the decrease of D. The promotion of propulsion performance corresponds to this result. It has been known that the higher discharge power produces a larger arc current, so as to produce a better constriction effect and a higher density of the plasma. Therefore, with the increase of discharge power, the promotion effect of the impulse bit for the multi-anode VAT is more remarkable than that of the conventional VAT.

The thrust-to-power ratio also exhibits distinct differences between the conventional VAT and the multi-anode VATs, which are shown in figure 12(b). The thrust-to-power ratio of the conventional VAT declines slightly with the increase of capacitance energy, from 0.59 at 4 J to 0.40 at 9.8 J. While, the multi-anode VAT shows an evident upward trend of thrust-to-power ratio with the increase of discharge power. The upward trend of thrust-to-power ratio is more significant with the decrease of D. The multi-anode VAT with D of 100 mm has a lower thrust-to-power ratio than that of the conventional VAT when the discharge power is small. However, the thrust-to-power ratio of the multi-anode VATs with D of 40 mm is obviously greater than that of the conventional VAT. It increases to 1.81 at capacitance energy of 9.8 J, which is 4.5 times that of the conventional VAT. The thrust-to-power ratio could reflect the propulsion performance per unit of energy consumed. The reason for the decline of thrust-to-power ratio of the conventional VAT with higher discharge power is that the greater discharge current will promote the generation of plasma, accompanied by the increase of the plasma diffusing in radial direction. The energy lost with the diffusing plasma increases synchronously, which causes the decrease of thrust-to-power ratio. In previous research, Lun [31] obtained a similar result with the total cumulative pulse charges as the variable. However, the higher arc current in the multi-anode VAT will exert a greater pinch force on the plasma plume, therefore the increase of the diffusion is inhibited. The decrease of D also contributes to the reduction of plasma diffusion in radial direction. As a result, the increase of lost energy is suppressed and the energy utilization of the multi-anode VAT is promoted.

Consequently, the propulsion performance of the multi-anode VAT have been verified to be superior to the conventional VAT. It has the possibility to be applied in the field of nanosatellite propulsion.

## 6. Conclusion and prospect

In this paper, a multi-anode electrode geometry is proposed to realize the radial constriction of the plasma and enhance the plasma plume intensity in axial direction. The VAT based on the multi-anode electrode geometry has a better propulsion performance in the same discharge parameters compared with the conventional VAT. The innovation is mainly embodied in the following aspects:

Firstly, the multi-anode electrode geometry has a different discharge characteristics from the conventional electrode geometry. It can direct the current to flow through the plasma plume, which is conducive to the formation of directional plasma plume.

Secondly, the multi-anode electrode geometry is able to reduce the radial diffusion of plasma and increase the plasma utilization of the VAT.

Thirdly, the multi-anode VAT is able to reduce the loss of discharge energy compared with the conventional VAT. The advantage is further promoted with the increase of discharge power.

This paper focuses on the discharge characteristics and plasma propagation characteristics of the multi-anode electrode geometry, and initially proves its advantage in the application of VAT. While, in order to fully understand the plasma generation and propagation mechanism of this electrode geometry, the parameters of multi-anode electrode geometry need to be discussed comprehensively. Limited to the length of the article and the current study, these issues will be discussed in the future study.

**Acknowledge**

This work was supported by the National Natural Science Foundation of China (No. 51577011) and the Graduate Innovation Project of Beijing Jiaotong University (No. 2016YJS147).